\documentclass[12pt]{article}
\pdfoutput=1
\usepackage[utf8]{inputenc}
\usepackage{amsmath,amsfonts,amssymb}
\usepackage{jheppub}

\usepackage{enumerate}
\usepackage{slashed}

\newcommand{\be}{\begin{equation}}
\newcommand{\ee}{\end{equation}}
\newcommand{\bea}{\begin{equation}\begin{aligned}}
\newcommand{\eea}{\end{aligned}\end{equation}}
\newcommand{\ba}{\begin{array}}
\newcommand{\ea}{\end{array}}

\newcommand{\M}{{\mathfrak M}}
\newcommand{\CN}{{\mathcal N}}
\newcommand{\CO}{{\mathcal O}}

\newcommand{\CW}{{\mathcal W}}

\newcommand\pd{p^\dag}
\newcommand\qd{q^{\dagger}}

\newcommand\Qd{Q^{\dagger}}
\newcommand\Pd{P^{\dagger}}

\def\a{\alpha} 
\def\b{\beta}

\newcommand{\cC}{\mathcal{C}}

\newcommand{\cL}{\mathcal{L}}

\newcommand{\cN}{\mathcal{N}}
\newcommand{\cO}{\mathcal{O}}

\newcommand{\cW}{\mathcal{W}}

\newcommand{\bR}{\mathbb{R}}

\newcommand{\bZ}{\mathbb{Z}}
\newcommand{\sT}{\mathsf{T}}

\newcommand{\wb}{\overline}
\newcommand{\smat}[1]{\big( \begin{smallmatrix} #1 \end{smallmatrix} \big)}
\newcommand{\ie}{\textit{i.e.}}
\newcommand{\matht}[1]{\ensuremath{\boldsymbol{#1}}}

\newcommand{\tp}{\mathrel{\widetilde+}}
\DeclareMathOperator{\Tr}{Tr}

\title{\matht{\mathcal{N}=1} dualities in 2+1 dimensions}

\preprint{SISSA 08/2018/FISI}

\author{Francesco Benini, }
\author{Sergio Benvenuti}

\affiliation{International School of Advanced Studies (SISSA), Via Bonomea 265, 34136 Trieste, Italy}
\affiliation{INFN, Sezione di Trieste, Via Valerio 2, 34127 Trieste, Italy}
\emailAdd{fbenini@sissa.it, benve79@gmail.com}

\abstract{We consider minimally supersymmetric QCD in 2+1 dimensions, with Chern-Simons and superpotential interactions. We propose an infrared $SU(N) \leftrightarrow U(k)$ duality involving gauge-singlet fields on one of the two sides. It shares qualitative features both with $3d$ bosonization and with $4d$ Seiberg duality. We provide a few consistency checks of the proposal, mapping the structure of vacua and performing perturbative computations in the $\varepsilon$-expansion.}

\begin{document}
\maketitle

\section{Introduction and results}

The last couple of years saw the discovery of three-dimensional bosonization: infrared (IR) dualities enjoyed by gauge theories with Chern-Simons interactions \cite{Aharony:2011jz, Giombi:2011kc, Barkeshli:2014ida, Aharony:2012nh, Jain:2013gza, Son:2015xqa, Aharony:2015mjs, Karch:2016sxi, Murugan:2016zal, Seiberg:2016gmd, Hsin:2016blu, Karch:2016aux, Metlitski:2016dht, Aharony:2016jvv, Benini:2017dus, Komargodski:2017keh, Gomis:2017ixy, Armoni:2017jkl, Cordova:2017vab, Benini:2017aed, Jensen:2017bjo, Cordova:2017kue}, similar in spirit to the particle-vortex duality \cite{Peskin:1977kp, Dasgupta:1981zz} and to dualities with extended $\CN\geq 2$ supersymmetry studied for more than two decades \cite{Intriligator:1996ex, Aharony:1997bx, Aharony:1997gp, Giveon:2008zn, Benini:2011mf}.

In this paper we want to tackle the case of $3d$ minimal supersymmetry, namely $\CN=1$ (see \cite{Gates:1983nr}), an interesting bridge between $\CN=0$ and $\CN \geq 2$ for many reasons.

With $3d$ $\CN=1$ supersymmetry all supermultiplets (except for those containing conserved currents) are \emph{long}: there is no protected sector analogous to the chiral ring. Moreover, the interactions are not protected. This makes the analysis of $\cN=1$ theories similar to the $\CN=0$ case, and the dualities quite powerful: usually for $\CN \geq 2$ dualities, it is only known how to map the protected operators to the dual theory. On the other hand if we have an $\CN=1$ duality and we know the mapping of a supermultiplet, we can, in principle, deform the $\CN=1$ duality to an $\CN=0$ duality.

See \cite{Gremm:1999su, Ohta:1997fr, Kitao:1998mf, Bergman:1999na, Gukov:2002es,Ooguri:2008dk, Armoni:2009vv,Forcella:2009jj,Amariti:2014ewa} for earlier work on $3d$ $\CN=1$ gauge theories and dualities.

Let us also mention that there might be experimental realizations of $(2{+}1)$-di\-men\-sio\-nal systems with low amount of supersymmetry. This is due to the phenomenon of \emph{emergent supersymmetry} \cite{Lee:2006if, Grover:2013rc, Jian:2014pca, Zerf:2016fti,Fei:2016sgs,Jian:2016zll, Li:2017dkj}. Roughly speaking, if the massless matter of some model is supersymmetric, then also the interactions of the infrared  fixed point may be supersymmetric: the renormalization group flow may land on the SUSY fixed point.

\subsection*{\matht{\CN=1 \;\; SU(N) \leftrightarrow U(k)} duality with gauge singlets}

In this paper we focus on $3d$ $\CN=1$ models with unitary gauge groups and fundamental flavors. We are interested in a supersymmetric analog of the non-Abelian bosonization. Bosonization maps \emph{critical scalars} (\ie{} with quartic $\phi^4$ interactions)   to \emph{regular fermions}. In a supersymmetric theory, we expect critical scalars to be paired with \emph{critical fermions} (\ie{} with interactions similar to the Gross-Neveu-Yukawa model), and regular fermions to be paired with \emph{regular scalars}. A supersymmetric $U(N_c)$ or $SU(N_c)$ gauge theory with a matter multiplet $\Phi$ in the fundamental representation does not admit interactions of the form $\phi^4$: the generic superpotential $\cW = |\Phi|^4$ leads to interactions of the form $\phi^6 + \phi^2\psi\psi$, so we have \emph{regular matter}. In order to have an $\CN=1$ duality mapping \emph{critical matter} to \emph{regular matter}, we expect it is necessary to introduce additional fields, the simplest option being gauge-singlet supermultiplets.

In the case of a single flavor, we introduce one gauge-singlet real superfield $H$ and propose the following duality:
\be
\label{basicproposal}
\ba{c}U(k)_{N+\frac{k}{2}-\frac{1}{2},\, N - \frac{1}{2}} \\[.4em]
\text{with 1 flavor $Q$} \\[.4em]
\CW=  - \frac14 \big( \sum_{i=1}^k Q_iQ_i^\dag \big)^2 \ea
\qquad\longleftrightarrow\qquad 
\ba{c} \text{$SU(N)_{-k-\frac N2+\frac12}$ with 1 flavor $P$} \\[.4em] 
\text{and a gauge-singlet $H$} \\[.4em]
\CW= H \, \sum_{i=1}^N P_i\Pd_i - \frac13 H^3 \;. \ea 
\ee
The sign of the superpotential on the left-hand side is important: the physics would be different with the other sign. Notice that a parity transformation changes the sign of both the Chern-Simons terms and the superpotential, while the relative sign remains unchanged. On the right-hand side, one can redefine $H \to -H$ and change the overall sign of the superpotential, while the relative sign between the two terms is important. On the $U(k)$ side we have \emph{regular matter}, on the $SU(N)$ side we have \emph{critical matter}.

The ``topological" global symmetry on the $U(k)$ side is mapped to the baryonic symmetry of the $SU(N)$ gauge theory. Accordingly, monopole operators are mapped to baryonic operators. This is similar to what happens in the non-supersymmetric bosonization dualities.

As for the operators in the sector with vanishing global $U(1)$ charge, our proposal is the that the meson is mapped to the singlet:
\be
\label{basicmap}
Q\Qd \qquad\longleftrightarrow\qquad H \;.
\ee
This is analogous to the mapping in $4d$ $\CN=1$ Seiberg duality \cite{Seiberg:1994pq} or $3d$ Aharony dualities \cite{Aharony:1997gp}.%
\footnote{Notice however that here the operators $Q\Qd$ and $H$ are not protected by supersymmetry: they are \emph{long} supermultiplets.} 

Deforming the two theories with a superpotential term $m\, Q\Qd \leftrightarrow m \, H$, the vacua display an interesting behavior: for $m \leq 0$ there is one vacuum, while for $m>0$ there are two isolated vacua.%
\footnote{For $N=1$ the structure is a bit different. We study the case $N=1$ in detail in Section~\ref{sec: abelian}.}
Each vacuum is gapped and hosts a certain topological sector. Since the same vacuum structure and TQFTs appear on both sides of the duality, we have a first consistency check of our proposal.

A non-trivial implication of the duality \eqref{basicproposal} and of \eqref{basicmap} is that $(Q\Qd)^2$ is mapped to $H^2$. Since $(Q\Qd)^2$ is in the superpotential, in the IR it must be the case that the dimension $\Delta\big[ (Q\Qd)^2\big]_{IR}>2$, and so $\Delta\big[\int d^2\theta (Q\Qd)^2\big]_{IR}>3$. In order for the duality to be correct, then it must also be the case that on the $SU(N)$ side  $\Delta[H^2]_{IR}>2$. This is a somewhat surprising statement, since $H^2$ is a mass term, with $\Delta[H^2]_{UV}=1$. 

In order to gain further insight, we performed a perturbative computation in the ungauged model (cubic Wess-Zumino model) with superpotential
\be
\label{WZM}
\CW= H \sum_{i=1}^N P_i\Pd_i  - H^3
\ee
at two loops in the $4-\varepsilon$ expansion (with Pade resummation). The result is that indeed in the Wess-Zumino model $\Delta[H^2]_{IR}>2$ if $N\geq 1$.  Gauging the $SU(N)$ or $U(N)$ global symmetry of the Wess-Zumino model, for large enough Chern-Simon level, does not spoil the inequality $\Delta[H^2]_{IR}>2$. The fact that $H^2$ is irrelevant in the IR is a consistency check of our proposed duality, and justifies our choice of superpotential in \eqref{basicproposal} which does not include an $H^2$ term. We expect this picture and the duality \eqref{basicproposal} to be correct for any $N \geq 1$ and $k \geq 1$.

There are related but different versions of the duality \eqref{basicproposal}. One example\footnote{One can go from \eqref{basicproposal} to (the time reversal of) \eqref{basicproposal2} either by gauging the global symmetry and renaming $k \leftrightarrow N$, or by ``flipping" the operators $Q\Qd \leftrightarrow H$ in \eqref{basicproposal}.} is 
\be
\label{basicproposal2}
\ba{c} U(k)_{-N-\frac{k}{2}+\frac{1}{2},\, -N+\frac{1}{2}} \\[.4em]
\text{with 1 flavor $Q$ and a singlet $H$}  \\[.4em]
\CW=  H \sum_{i=1}^k Q_i\Qd_i - \frac13 H^3 \ea 
\qquad\longleftrightarrow\qquad 
\ba{c} \text{$SU(N)_{k + \frac N2-\frac12}$ with 1 flavor $P$} \\[.4em]
\CW= - \frac14 \big( \sum_{i=1}^N P_iP_i^\dag \big)^2 \;. \ea 
\ee
 This version is very similar to \eqref{basicproposal}: the superpotential contains $H^3$ but not $H^2$ and upon mass deforming the theories, there are two vacua merging into a single vacuum. Now on the $U(k)$ side we have \emph{critical matter}, on the $SU(N)$ side we have \emph{regular matter}. The duality \eqref{basicproposal2} is expected to be valid for $k \geq 1$ and $N>1$.

However, \eqref{basicproposal2} cannot be valid for $N=1$: in this case the $SU(N)$ side becomes a free complex superfield, that displays enhanced $\CN=2$ supersymmetry, has a non-trivial moduli space of vacua and a different vacuum structure upon mass deformations (a single vacuum for both signs of the mass). Our proposal in this case is\footnote{Notice that for $k=1$ the gauge theory $U(1)_{-\frac 12}$ displays explicit $\CN=2$ supersymmetry, and we recover the well-known $\CN=2$ basic mirror symmetry.}
\be
\label{basicproposal2p}
\ba{c} U(k)_{-\frac{k+1}{2},\, -\frac{1}{2}} \\[.4em]
\text{with 1 flavor $Q$ and a singlet $\tilde{H}$}  \\[.4em]
\CW=  \tilde{H} \sum_{i=1}^k Q_i\Qd_i - \frac12 \tilde{H}^2 \ea 
\qquad\longleftrightarrow\qquad 
\ba{c} \text{Free $\CN=2$}\\ \text{chiral multiplet $P$} \;. \ea 
\ee
In this way one-loop effects balance the superpotential term $\tilde H^2$ and allow a non-trivial moduli space of vacua for the $U(k)$ theory. Also the vacuum structure upon mass deformations is the same.

Summing up, our proposal is the following: the dual of an $\CN=1$ $SU/U$ gauge theory with \emph{regular matter} (and hence no singlets) is a $U/SU$ theory with \emph{critical matter} (so there are singlet fields). The superpotential term for the singlet fields is cubic. The only exception is when the \emph{regular matter} is actually free: then the qualitative structure is different and there is a quadratic superpotential term for the singlet field on the dual side.

\subsection*{Further directions}
As for the non-Abelian dualities, it would be nice to investigate what happens outside their range of validity (especially in the generalized case with arbitrary number of flavors $N_f>1$ discussed in Section \ref{sec: nflarge}), and study possible quantum phases as in \cite{Komargodski:2017keh, Gomis:2017ixy}.

It would also be interesting to study non-supersymmetric deformations and possibly make contact with the bosonization dualities and/or the non-supersymmetric dualities with bosons and fermions studied in \cite{Jain:2013gza, Benini:2017aed,Jensen:2017bjo}.  From the mapping of low-lying operators in terms of $\CN=1$ superfields that we provide, it is possible to read off the mapping of non-supersymmetric deformations. In particular notice that even if the top component of the superfield $H^2$ is an irrelevant SUSY deformation, the bottom component of $H^2$ is a relevant non-SUSY deformation.

In this paper we have analyzed the Wess-Zumino model \eqref{WZM} at two-loops in the \mbox{$\varepsilon$-expansion}. It would be interesting to increase the precision, either going to higher loops and possibly interpolating to a solvable $2d$ model \cite{Zerf:2016fti,Fei:2016sgs,Baggio:2017mas}, or employing the numerical bootstrap \cite{Bashkirov:2013vya, Bobev:2015vsa, Bobev:2015jxa}.

The Abelian duality we discuss in Section \ref{sec: abelian}  has a simple Type IIB brane description \textit{\`a la} Hanany-Witten, involving a D3-brane stretching between an NS5-brane and the simplest $pq$-web, or the S-dual configuration. The setup is similar to \cite{Gremm:1999su, Ohta:1997fr, Kitao:1998mf, Bergman:1999na, Gukov:2002es, Armoni:2009vv}. It would be nice to find a brane description also for the non-Abelian cases.

Finally, let us mention that the dualities of this paper might be useful to understand walls and boundary conditions of $4d$ $\CN=1$ SQCD \cite{Acharya:2001dz, Armoni:2009vv,Gaiotto:2017tne,Gomis:2017ixy}.

\subsection*{Organization of the paper}

In Section \ref{sec: abelian} we study in detail the Abelian case $N=k=1$. In this case the duality \eqref{basicproposal} relates $\CN=1$ SQED with one flavor and a quartic superpotential on one side, and a cubic Wess-Zumino model on the other side. The vacua form a circle for $m<0$, while there are two isolated vacua for $m>0$. The  duality \eqref{basicproposal2} instead becomes the well-known duality between $\CN=2$ $U(1)_{\frac12}$ SQED with one flavor and a free chiral field. Both cases are parity-invariant, a symmetry which on the SQED side emerges in the IR. We also find two additional $U(1) \leftrightarrow U(1)$ dualities relating regular to critical matter.

In Section \ref{Nonabelian} we discuss the non-Abelian case. We present four different versions of the duality, involving in various ways gauge groups $SU(N_c)$ and $U(N_c)$. The various dual pairs can be related to each other by gauging a $U(1)$ global symmetry. Moreover, we state the natural conjectures for number of flavors greater than one.

In Appendix \ref{loops} we discuss some details of the perturbative computations for the cubic $\cN=1$ Wess-Zumino model. Appendix \ref{app: level-rank} provides the topological level-rank dualities in $\CN=1$ notation.

\vspace{.7cm}

\paragraph{Note added.}
When this work was under completion, we received \cite{Bashmakov:2018wts} which, among other things, discusses $3d$ $SU/U(N)$ $\CN=1$ gauge theories with a fundamental flavor. The dualities studied  in \cite{Bashmakov:2018wts} are similar to ours, but have different superpotential interactions and do not involve gauge-singlet fields.


\section{\matht{\CN=1} SQED with \matht{N_f=1} and its Wess-Zumino dual}
\label{sec: abelian}

In this Section we discuss in detail the case $N=k=1$ of the proposed dualities \eqref{basicproposal}. 

We use the $3d$ $\CN=1$ superfields of \cite{Gates:1983nr}, that in the spin-0 case expand as
\be
A(\theta) = a + \theta \lambda_{A} + \theta^2 F_{A} \;.
\ee
We use upper-case letters to denote the whole superfield and lower-case letters to denote the bottom component. The Lagrangians have standard kinetic terms, and the interactions are encoded in a real superpotential $\CW=\CW(A_i)$.

One important remark is that $d^2\theta$ is parity-odd. Therefore a gauge theory with zero Chern-Simon level is parity invariant if it is possible to assign parity quantum numbers to the matter superfields such that $\CW$ is parity-odd.

\subsection[Warm-up: the $\CN=2$ duality]{Warm-up: the \matht{\CN=2} duality}
\label{sec: N=2 warm up}

It is instructive to start with the basic $\cN=2$ duality between the Chern-Simons gauge theory $U(1)_{1/2}$ with 1 flavor $Q$ and a free chiral field $P$. Written in $\cN=1$ notation this is
\be
\label{N=2 duality}
\ba{c} \text{$U(1)_{\frac12}$ with 1 flavor $Q$ and a singlet $\Psi$} \\[.5em]
\CW= \Psi Q \Qd -\frac{1}{2}\Psi^2
\ea 
\quad\longleftrightarrow\quad
\ba{c}
\text{Free complex superfield $P$} \\[.5em]
\CW= 0 \;.
\ea
\ee
Here $\Psi$ is a real scalar superfield, that completes the $\cN=1$ vector multiplet to the $\cN=2$ vector multiplet. Note that one cannot integrate out the field $\Psi$, since the coefficient in front of $\Psi^2$ is not parametrically large with respect to the scale of the gauge coupling: it is fixed in terms of the CS level.

The free complex superfield $P$ has a $U(1)$ symmetry that rotates it. This corresponds to the magnetic (or topological) $U(1)_M$ symmetry on the left-hand side (LHS).
The right-hand side (RHS) is manifestly time-reversal invariant; instead on the LHS time reversal is an emergent symmetry in the infrared.

The operators, collected into supefields, are mapped according to
\be
\label{N=2 duality operators}
\left\{ \ba{c} \M \\  \Psi  \ea  \right\} \qquad\longleftrightarrow\qquad \left\{ \ba{c} P \\  P \Pd  \ea  \right\} \;.
\ee
Here $\M$ is the gauge-invariant supersymmetric monopole. We used that the bottom component $\psi$ of $\Psi(\theta) = \psi + \theta \lambda_{\Psi} + \theta^2 F_{\Psi}$ is in the same $\CN=2$ supermultiplet as the magnetic $U(1)_M$ current, and has dimension $\Delta[\psi] =1$. From the superpotential we can instead find the top component of the superfield $\Psi$:
\be
\label{bottom component}
F_\Psi = q\qd -\psi
\ee
where $q$ is the bottom component of $Q(\theta)$. Thus a particular linear combination of $q\qd$ and $\psi$ is a supersymmetry descendant of $\psi$, and has $\Delta[q\qd-\psi]=2$. Another linear combination---simply $\psi$---has $\Delta[\psi]=1$. We can now take the top component (\ref{bottom component}) and use it as the bottom component of a new superfield. This allows us to write the map of superfields
\be
Q\Qd-\Psi \qquad\longleftrightarrow\qquad D_\a P \, D^\a \Pd
\ee
where $D_\alpha$ is the superderivative. This could be used to infer new relations between their components, and so on.

Notice also that the duality implies the quantum relation
\be
\M \, \M^\dagger = \Psi
\ee
in the gauge theory.

\paragraph{SUSY deformations.} The free complex superfield $P$ has a single $\cN=1$ relevant deformation compatible with the $U(1)$ symmetry: a superpotential mass term $\delta \cW = m PP^\dag$ with $m\in \bR$. In $\cN=2$ notation, this is a ``real mass''. The quartic superpotential deformation $\delta \cW = \frac14 \alpha PP^\dag PP^\dag$, that gives a sextic potential $V = \alpha^2 |p|^6$, is marginally irrelevant.

Therefore, one interesting consequence of the duality is that, on the LHS, the superpotential deformations $\Psi^2$ and $Q\Qd$ are \emph{irrelevant} (more precisely, marginally irrelevant since $\Delta_{IR}=2$) in the infrared CFT, even though they are clearly relevant in the ultraviolet where they have $\Delta_{UV}=1$. In other words, in the UV there are $3$ relevant global-symmetry-invariant deformations: $\Psi, \Psi^2, Q\Qd$. In the IR only one of them ($\Psi$) is relevant.

The relevant $\CN=1$ deformation of the IR CFT, invariant under the global $U(1)$ symmetry, is thus 
\be
\label{defN=2duality}
\delta \CW = m\, \Psi \qquad\longleftrightarrow\qquad \delta \CW = m \ P \Pd \;.
\ee
As we said, this is actually the $\CN=2$ preserving ``real mass" deformation, and it breaks parity invariance. The free theory of $P$ has, obviously, a single gapped vacuum both for $m>0$ and $m<0$.

Let us quickly analyze the phases on the SQED side. For $m>0$ there is a vacuum where $\psi$ gets a positive VEV and $Q$ gets positive mass. Integrating it out, the CS level is shifted to $U(1)_1$ and the quadratic superpotential term is shifted to $-\Psi^2$ by a one-loop effect. The F-term equation for the effective superpotential $\cW_\text{eff} = m\Psi - \Psi^2$ is consistently solved by $\psi = m/2$, $q=0$ and the vacuum is gapped. For $m<0$ there is a vacuum where $|q|^2 = |m|$ and $\psi=0$: the Higgs mechanism takes place and the vacuum is gapped. Therefore the phases match.

For $m<0$ we could have considered the possibility that $\psi$ gets a negative VEV and $Q$ gets negative mass. Integrating it out, the CS level is shifted to $U(1)_0$ and the quadratic superpotential term is shifted to zero by a one-loop effect. The F-term equation for the effective superpotential $\cW_\text{eff} = m\Psi$ does not allow for a VEV of $\psi$, leading to a contradiction. Notice also that since $P$ is a free field, $PP^\dag$ is a positive operator: if the duality is correct, $\Psi$ should not be able to get a negative VEV.

\subsection[The basic $\CN=1$ duality]{The basic \matht{\CN=1} duality}

From the $\cN=2$ duality and exploiting the operator map \eqref{N=2 duality operators}, we can obtain a genuine $\CN=1$ duality. We ``flip" the real $\CN=1$ superfield $\Psi \leftrightarrow P \Pd$ on both sides, \ie{} we introduce a parity-odd real superfield $H$ and we couple it through the interactions 
\be
\delta \CW = H \, \Psi \qquad\longleftrightarrow\qquad  \delta\CW = H \, P \Pd
\ee
with large coefficient. Notice that this deformation is relevant on both sides of the $\cN=2$ duality.

On the LHS both $\Psi$ and $H$ become massive and can be integrated out. This generates quartic superpotential interactions for the flavor $Q$, and the coefficient will be renormalized to a critical value. On the RHS we obtain an interacting Wess-Zumino (WZ) model. No symmetry prevents a superpotential term $H^3$ to be generated by quantum effects. In this way we obtain a new genuinely $\cN=1$ duality
\be
\label{basic N=1 duality}
\ba{c} \text{$U(1)_{\frac12}$ with 1 flavor $Q$} \\[.5em]
\CW= - \frac14  Q \Qd Q \Qd \ea 
\qquad\longleftrightarrow\qquad
\ba{c} \text{WZ model with $P$, $H$} \\[.5em]
\CW=  H P \Pd - \frac{1}{3} H^3 \;.
\ea 
\ee
Here $Q$ and $P$ are complex superfields, while $H$ is a real superfield. In the WZ model there is also an extra gravitational coupling $-\mathrm{CS_g}$.

On the LHS we tuned the term $Q\Qd$ to zero. The coefficient $-\frac14$ was chosen for later convenience, but its sign is physical and important. The WZ model on the RHS is manifestly parity invariant ($H$ is parity-odd). For this reason the terms $H^2$ and $P\Pd$ are not generated. We tuned the term $H$ to zero. Since we can redefine $H \rightarrow -H$, only the relative sign of the couplings in front of $H P\Pd$ and $H^3$ has physical meaning. Performing loop computations in the $\varepsilon$-expansion, we will confirm that the term $H^3$ is generated, and we will see that $H P\Pd$ and $H^3$ have opposite sign at the RG fixed point.

The basic operator map is
\be
\left\{ \ba{c} \M \\ Q\Qd   \ea  \right\} \qquad\longleftrightarrow\qquad \left\{ \ba{c} P \\ H \ea  \right\} \;.
\ee
Similarly to above, supersymmetry imposes some relations. We focus on the operators neutral under the $U(1)$ global symmetry. The top components of the fundamental superfields on the RHS are
\be
F_P = 2 h p \;, \qquad\qquad F_{H} = p\pd - h^2 \;.
\ee
Thus the operator $p\pd - h^2$ is a supersymmetry descendant of $h$. Another linear combination of $p\pd$ and $h^2$, which we formally denote as $p\pd \tp h^2$, is instead a superconformal primary. (Moreover, a linear combination of $h p \pd$ and $\wb\lambda_P \lambda_P$ is a descendant of $p\pd$.)

On the LHS of the duality we have
\be
F_Q = - q(q\qd) \;.
\ee 
The $\theta^2$-component of the superfield $Q\Qd$ is $\wb\lambda_Q\lambda_Q + qF_Q^\dag + \qd F_Q$. Hence one linear combination of $\wb\lambda_Q \lambda_Q$ and $(q\qd)^2$ is a descendant of $q\qd$, while another combination, which we formally denote as $\wb\lambda_Q \lambda_Q \tp (q\qd)^2$, is a superconformal primary (the precise coefficients in front of the two terms depend on the computation scheme used). We thus have one more operator mapping:
\be
\wb\lambda_Q \lambda_Q \tp (q\qd)^2 \qquad\longleftrightarrow\qquad p \pd \tp h^2 \;,
\ee
which can be upgraded to a map between supermultiplets:
\be
D_\a Q \, D^\a \Qd \tp (Q\Qd)^2 \qquad\longleftrightarrow\qquad P\Pd \tp H^2\;.
\ee

\paragraph{SUSY deformations.} From the last operator mapping, one crucial feature of our proposed scenario follows. On the SQED side, we know that the primary operator  $D_\a Q D^\a \Qd \tp (Q\Qd)^2$ becomes in the IR an irrelevant superpotential deformation, since it already appears in the Lagrangian. So, if the duality is correct, it must be the case that on the RHS the operator $\cO_s = P\Pd \tp H^2$ satisifes $\Delta[\cO_s]\geq 2$ and is an irrelevant superpotential deformation---even though $\Delta[\cO_s]_{UV}=1$ in the free UV theory. This feature is very similar to what happens in the $\CN=2$ duality studied above. In order to test this proposal, we compute $\Delta[\cO_s]$ in the $\varepsilon$-expansion in the next subsection.

Let us stress that it is essential, for the duality to work, to make sure that there are no relevant deformations corresponding to the superpotential terms $P\Pd$ and $H^2$. The WZ model on the RHS is time-reversal invariant in the UV, therefore those deformations will not be generated in any case. The theory on the LHS, instead, develops time-reversal invariance only in the IR. If the deformations $P\Pd$ or $H^2$ were relevant, they would be activated in the IR and the SQED theory on the LHS could not hit the time-reversal invariant fixed point (one would need extra tuning, which however is not available in the SQED theory in the UV).

\subsection[IR irrelevance of the``mass term" $\cO_s = |P|^2 \tp H^2$ in the $\varepsilon$-expansion]{IR irrelevance of the``mass term" \matht{\cO_s = |P|^2 \tp H^2} in the \matht{\varepsilon}-expansion}
\label{2loop}

We consider the cubic Wess-Zumino model with $\CN=1$ supersymmetry and superpotential
\be
\CW = \frac{g_2}{2}  H P \Pd + \frac{g_3}{6}  H^3 \;.
\ee
Working in the $4-\epsilon$ expansion, we have computed (see Appendix \ref{loops}) the two-loop beta-functions for the model. The numerical values of the couplings at the physically sensible fixed point are
\bea
\frac{g_2}{4\pi \sqrt{\varepsilon}} &= 0.38237 + 0.06895 \varepsilon + O(\varepsilon^2) \\
-\frac{g_3}{4\pi\sqrt{\varepsilon}} &= 0.41439 + 0.07202 \varepsilon  + O(\varepsilon^2) \;.
\eea
Notice in particular that the couplings in front of $ H P \Pd$ and $ H^3$ have opposite sign.

The two-loop scaling dimensions of the elementary fields are
\bea
\Delta[H] &= 1- 0.26793 \varepsilon -0.000028\varepsilon^2+O(\varepsilon^3) &&\simeq_\text{Pade}\; 0.732 \\
\Delta[P] &= 1- 0.35379 \varepsilon -0.00258\varepsilon^2+O(\varepsilon^3)  &&\simeq_\text{Pade}\; 0.644 \;.
\eea
On the right we quoted the Pade$[1,1]$ resummed value at $\varepsilon=1$.

The quadratic operators in the symmetric-traceless representation of the $O(2)$ global symmetry have scaling dimension
\be
\Delta[P^2] = 2 - 0.41517 \varepsilon -0.00887\varepsilon^2+O(\varepsilon^3)  \;\;\simeq_\text{Pade}\; 1.472 \;.
\ee
There are two quadratic singlets under the $O(2)$ global symmetry. One is $g_2 P\Pd + g_3  H^2$: this is a supersymmetry descendant of $H$ therefore its scaling dimensions is $\Delta[H]+1$ at the IR fixed point.

The other singlet operator is the superconformal primary $\cO_s$. Its precise form in our computation scheme and at two-loops is 
\be
\cO_s = \left(1.845466+ 2.061069 \varepsilon + O(\varepsilon^2)\right) H^2 + P\Pd \;.
\ee
Its scaling dimension is
\be
\Delta[\CO_s] = 2 + 0.12448 \varepsilon -0.13902\varepsilon^2+O(\varepsilon^3) \;\; \simeq_\text{Pade}\; 2.058 \;.
\ee
We see that $\Delta[\CO_s] > 2$, as required by our proposed duality.

On the other hand, $\Delta[P], \Delta[P^2]<2$ so the two superpotential monopole deformations $\M+\M^\dagger$ and $\M^2+(\M^2)^\dagger$ (which break the $U(1)$ topological symmetry completely and to $\mathbb{Z}_2$, respectively) are relevant deformations in the IR CFT.%
\footnote{As described in \cite{Benini:2017dud} for gauge theories with $\CN=2$ supersymmetry, it is nevertheless possible that the theory $\CN=1$ SQED with superpotential $\CW=\M^h + (\M^h)^\dagger$ exists also for $h>2$, even if it is not reachable with an RG flow starting from SQED with no monopole superpotential. It would be nice to investigate this issue further.}

\subsection{Relevant deformations and vacua}

Having established that there is only one deformation that preserves $\CN=1$ supersymmetry and the $U(1)\rtimes\bZ_2^\cC$ global symmetry (where $\bZ_2^\cC$ is charge conjugation), namely
\be
\frac m2 \, Q\Qd \qquad\longleftrightarrow\qquad   m \, H \;,
\ee
we now proceed to study the different phases that one obtains when turning on such a deformation, in the two dual theories respectively. We find that for $m>0$ there are two isolated gapped vacua (corresponding to the broken IR time-reversal symmetry), while for $m<0$ there is an $S^1$ Goldstone boson with a free fermion.%
\footnote{Very similar arguments work in the non-Abelian case, that we will study in Section \ref{Nonabelian}.}

As we will show below, it is convenient to keep track of counteterms for background fields, in particular for a gauge field $B$ that couples to the $U(1)$ global symmetry and for the metric. We use here the following notation for the various component fields. In the SQED theory the fields are a complex supermultiplet $Q = (q,\psi, F_Q)$, the gauge field $a$ and the gaugino $\lambda$. In the WZ model there is a real supermultiplet $H = (h, \eta, F_H)$ and a complex supermultiplet $P = (p, \chi, F_P)$. We recall that when integrating out a Majorana fermion with negative mass we generate the gravitational coupling $-\mathrm{CS_g}$, and that $U(1)_1$ is equivalent to $-2\mathrm{CS_g}$.

\paragraph{SQED side.} We study the theory with superpotential
\be
\cW = \frac m2 Q^\dag Q - \frac14 Q^\dag Q Q^\dag Q \;.
\ee
We also couple to the topological symmetry via $\frac1{2\pi} adB$. The F-term is
\be
F_Q = q\big( m - |q|^2 \big) \;.
\ee
This gives the following potential and fermionic interactions:
\bea
V &= |q|^2 \big( m^2 - 2m |q|^2 + |q|^4\big) \\
\cL_{\psi^2} &= \big( m- 2|q|^2\big) \wb\psi\psi - \frac12 \big( q^2 \wb\psi\psi^c + c.c. \big) + \big( i q \wb\psi \lambda + c.c. \big) - \wb\lambda \lambda \;.
\eea
It is useful to write fermionic interactions in a real notation.%
\footnote{Recall that Majorana fermions satisfy $\wb\psi_a^c = C \gamma_0^\sT \psi_a^* = \psi_a$. For Majorana fermions, $\wb\psi \chi = \wb\chi \psi$.}
Defining $Q = Q_1 + i Q_2$ in terms of real superfields $Q_a$, we find
\be
\cL_{\psi^2} = \big( m - 3 q_1^2 - q_2^2 \big) \wb\psi_1\psi_1 + \big( m - 3 q_2^2 - q_1^2 \big)\wb\psi_2 \psi_2 - 4q_1q_2\wb\psi_1\psi_2 + 2\big( q_1 \wb\psi_2\lambda - q_2 \wb\psi_1 \lambda \big) - \wb\lambda\lambda \;.
\ee
Depending on the sign of $m$ we find the following vacuum structure:
\begin{itemize}

\item $m>0$. There are two vacua.

One vacuum is at $q=0$ where $Q$ has mass $m$. Integrating it out, we get $\cN=1$ $U(1)_1$ CS. Since the gaugino is a free fermion with negative mass, it can be integrated out and we get $U(1)_1$ CS. This is a trivial gapped vacuum. If we keep into account the background counterterms, we have $-\frac1{4\pi} BdB -3\mathrm{CS_g}$.

The other vacuum is at $|q|^2 = m$. The gauge symmetry is Higgsed, both $\psi$ and $\lambda$ are massive, and we are left with a trivial gapped vacuum. More precisely, consider $q_1 = \sqrt m$ and $q_2=0$. We see that the radial superfield $\delta Q_1$ has mass $-2m$, the angular scalar $q_2$ participates in the Higgs mechanism, the angular fermion $\psi_2$ and $\lambda$ mix. The mass matrix is $\smat{0 & \sqrt m\\ \sqrt m& -1}$ whose eigenvalues are $\big( -1 \pm \sqrt{1+4m} \big)/2$, therefore one fermion has positive mass and one negative. We thus have $-2\mathrm{CS_g}$.

\item $m<0$. There is a vacuum at $q=0$ where $Q$ has mass $-|m|$. Integrating it out, we get $\cN=1$ $U(1)_0$ SYM, \ie{} a free massless fermion $\lambda$ and an $S^1$ free compact boson. We also have $-2\mathrm{CS_g}$.
\end{itemize}

\paragraph{Wess-Zumino side.} We study the deformed superpotential
\be
\cW = m H + H P \Pd - \frac13 H^3 \;.
\ee
Recall that we also have an extra gravitational coupling $-\mathrm{CS_g}$.
The superfield $P$ has charge 1 under the background gauge field $B$. The F-terms are
\be
F_H = m + |p|^2 - h^2 \;,\qquad\qquad F_P = 2hp \;.
\ee
They give potential and fermionic interactions
\bea
V &= \big( m + |p|^2 - h^2 \big)^2 + 4h^2 |p|^2 \\
\cL_{\psi^2} &= 2h \big( \wb\chi\chi - \wb \eta \eta \big)  + 2 p \wb\chi \eta + 2 p^\dag \wb\eta \chi \;.
\eea
Using the real notation $P = P_1 + i P_2$, the fermionic interactions are
\be
\cL_{\psi^2} = 2h (\wb\chi_a \chi_a - \wb\eta\eta) + 4 p_a \wb\chi_a \eta
\ee
with $a=1,2$. Depending on the sign of $m$ we find the following vacua:%
\footnote{Notice that when we integrate out $P$ with positive/negative mass, a negative/positive superpotential term $H^2$ is generated at one-loop. Because of the term $H^3$, though, the effect is negligible at large VEVs.}
\begin{itemize}
\item $m>0$. There are two vacua at $p=0$ and $h = \mp\sqrt m$, where $P$ has mass $\mp2\sqrt m$. Also $H$ is massive, with mass $\pm 2\sqrt m$ around its VEV. Integrating them out we are left with two trivial gapped vacua.

Notice that the VEV for $H$ breaks time-reversal symmetry, and the two vacua are related by that symmetry. Taking into account background counterterms, the vacuum with upper sign has $- \frac1{4\pi} BdB -3\mathrm{CS_g}$, the vacuum with lower sign has $-2\mathrm{CS_g}$.

\item $m<0$. There are vacua at $|p|^2 = |m|$ and $h=0$. The global $U(1)$ symmetry that rotates $P$ is spontaneously broken: we get a massless fermion and an $S^1$ free compact boson.

Using the real notation, we see that $H$ mixes with the radial part of $P$ around its VEV, giving two modes of opposite mass $\pm 2\sqrt{|m|}$. When we integrate them out we are left with $-2\mathrm{CS_g}$.

\end{itemize}

We see that the various phases perfectly match, including the counterterms for background fields.

\subsection{Other Abelian dualities}
\label{sec: gauging}

We can produce other Abelian dualities employing a gauging procedure. On both sides of the duality we add a CS term at level $\ell = 0$, $1$ or $-1$ for the background gauge field $B$, we couple it to a new background field $C$ and then make $B$ dynamical (we rename it $b$). In order to preserve $\cN=1$ supersymmetry, we should also introduce a real fermion $\gamma$ suitably coupled to the theory.

On the SQED side we have the Lagrangian terms
\be
\cL \supset \frac1{4\pi} ada + \frac1{2\pi} adb + \frac{\ell}{4\pi} bdb - \frac1{2\pi} bdC - \frac12 \wb\lambda\lambda - 2 \wb\lambda\gamma-\ell\, \wb\gamma\gamma \;.
\ee
Let us analyze the three cases in turn.

For $\ell=0$, the gauge field $b$ can be integrated out setting $a=C$. Integrating out $\gamma$ sets $\lambda=0$. We are left with an $\cN=2$ chiral multiplet $Q$ with charge $1$ under $C$ and a background counterterm $\frac1{4\pi} CdC$ (we neglect here gravitational couplings). In this theory the quartic superpotential interaction is marginally irrelevant and can be dropped.

For $\ell=1$, the gauge field $b$ can be integrated out, together with a massive fermionic eigenmode. We are left with
\be
\cL \supset \frac1{2\pi} adC - \frac1{4\pi} CdC \;.
\ee
The new theory has gauge group $U(1)_{-1/2}$.

For $\ell=-1$, the gauge field $b$ can be integrated out, leaving a bare CS level $2$, together with a massive fermionic eigenmode. We are left with
\be
\cL \supset \frac2{4\pi} ada - \frac1{2\pi} adC + \frac1{4\pi} CdC \;.
\ee
The new theory has gauge group $U(1)_{3/2}$.

On the WZ side we have the Lagrangian terms
\be
\cL \supset \frac{\ell}{4\pi} bdb - \frac1{2\pi} bdC - \big( \ell-\tfrac12\big)  \, \wb\gamma\gamma + i \big( p \wb\psi \gamma + c.c.\big) \;.
\ee
This is an Abelian gauge theory coupled to $P$ and $H$.

We thus find three new dualities.
\begin{itemize}
\item The case $\ell=0$ leads to
\be
\label{Abelian SU/U duality}
\ba{c} \text{Free complex field $Q$} \\[.5em]
\CW= 0 \ea 
\quad\longleftrightarrow\quad
\ba{c} \text{$U(1)_{-\frac12}$ with 1 flavor $P$ and a singlet $H$} \\[.5em]
\CW=  H P \Pd + \frac12 H^2 \;.
\ea 
\ee
On the LHS we dropped the quartic superpotential since it is marginally irrelevant. On the RHS, instead, we have used that in this particular case the superpotential deformation $H^2$ is relevant while $H^3$ becomes irrelevant. Indeed, the RHS is the $\cN=2$ SQED theory with one flavor. In this theory we know that $H^2$ is turned on at the fixed point, and we know that $H^3$ is irrelevant there. The gauging procedure allows then to go back from the $\cN=1$ duality to the $\cN=2$ duality.

\item In the case $\ell=1$ we find
\be
\label{Abelian U/U duality +}
\ba{c} \text{$U(1)_{-\frac12}$ with 1 flavor $Q$} \\[.5em]
\CW= - \frac14  Q \Qd Q \Qd \ea 
\quad\longleftrightarrow\quad
\ba{c} \text{$U(1)_\frac12$ with 1 flavor $P$ and singlet $H$} \\[.5em]
\CW=  H P \Pd - \frac{1}{3} H^3 \;. \ea 
\ee
The theory on the left is not the time-reversal of (\ref{basic N=1 duality}) because the CS level has opposite sign, but not the superpotential term. The duality suggests that the theory on the LHS has an $\cN=1$ fixed point, besides the $\cN=2$ fixed point. Even though both the $\cN=2$ (with $\cW = HPP^\dag- \frac12 H^2$) and the $\cN=1$ fixed point (with $\cW = HPP^\dag - \frac13 H^3$) are reachable from the same free UV model, let us emphasize that there are no $\cN=1$ RG flows going from one IR fixed point to the other one.

The phases of the two theories can be analyzed as before. For $m>0$ there is a gapped vacuum and an $S^1$ worth of vacua (parametrized by a Goldstone boson and with a free fermion). For $m<0$ there is a unique gapped vacuum.

\item In the case $\ell=-1$ we find
\be
\label{Abelian U/U duality -}
\ba{c} \text{$U(1)_\frac32$ with 1 flavor $Q$} \\[.5em]
\CW= - \frac14  Q \Qd Q \Qd \ea 
\quad\longleftrightarrow\quad
\ba{c} \text{$U(1)_{-\frac32}$ with 1 flavor $P$ and singlet $H$} \\[.5em]
\CW=  H P \Pd - \frac{1}{3} H^3 \;. \ea 
\ee
The phases are as follows. For $m>0$ there is a trivial gapped vacuum and a gapped vacuum with a topological theory $U(1)_2 \cong U(1)_{-2}$. For $m<0$ there is a trivial gapped vacuum.

\end{itemize}


\section{Non-Abelian dualities}\label{Nonabelian}

We propose that the Abelian $\cN=1$ dualities of the previous Section generalize to non-Abelian dualities. The four families of non-Abelian dualities are related by the gauging procedure described in Section \ref{sec: gauging}.

\subsection[$SU/U$ duality]{\matht{SU/U} duality}

The first family, that we call \matht{SU/U} and generalizes (\ref{Abelian SU/U duality}), is as follows. Theory A is
\be
\label{SU/U Theory A}
\ba{c} \text{$SU(N)_{k + \frac N2 - \frac12}$ with 1 flavor $Q$} \\[.5em] W = -\frac14 Q^\dag Q Q^\dag Q \;, \ea
\ee
while Theory B is
\be
\label{SU/U Theory B}
\ba{c} \text{$U(k)_{-N - \frac k2 + \frac12,\, -N+\frac12}$ with 1 flavor $ P$ and a real singlet $H$} \\[.7em] \CW = H  P^\dag  P - \frac13 H^3 \;. \ea
\ee
We consider this duality in the range $N \geq 2$, $k\geq 1$ (outside this range, it might still be possible to make sense of the duality along the lines of \cite{Komargodski:2017keh,Gomis:2017ixy}). The global symmetry is $O(2) = U(1) \rtimes \bZ_2^\cC$, where the second factor is charge conjugation.

In Theory B we can perform a field redefinition $H \to H + const.$ to remove a possible term $P^\dag P$ from the superpotential; we always assume we have removed such a  term. Next, the absence of a superpotential term $H^2$ is justified as follows. We start from the Wess-Zumino model with $k$ complex and $1$ real superfield, $\CW = H  P^\dag  P - \frac13 H^3$. In Appendix~\ref{loops} we show that for any $k>0$, the singlet quadratic operator of the form $\cO_s= P^\dag  P \tp H^2$ has $\Delta[\cO_s]>2$. Gauging the global $U(k)$ symmetry with a large enough Chern-Simons level will not spoil the relation $\Delta[\cO_s]>2$. So, for large enough Chern-Simons level, the correct superpotential does not contain the term $H^2$.

Finally, we have performed a tuning to zero of the mass term $Q^\dag Q$ in Theory A and of the linear term $H$ in Theory B. Those represent the only $\cN=1$ relevant deformation, identified on the two sides of the duality:
\be
\delta\cW = \frac m2\, Q^\dag Q \qquad\longleftrightarrow\qquad \delta\cW = m\, H \;.
\ee
We study the resulting phases below.

The case $N=1$ is special, because Theory A is a free $\cN=2$ chiral multiplet $Q$. In this case, the duality suggests that in Theory B---because of the particularly small CS level---the superpotential term $H^2$ is relevant and present in the theory. We propose that Theory B is
\be
\label{SU/U Theory B special N=1}
\ba{c} \text{$U(k)_{- \frac{k+1}2,\, -\frac12}$ with 1 flavor $ P$ and a real singlet $H$} \\[.7em] \CW = H  P^\dag  P + \frac 12 H^2 \ea
\ee
while Theory A is as in (\ref{SU/U Theory A}) with $N=1$ and $k\geq1$. For $k=1$ this is the $\cN=2$ $U(1)_{-\frac12}$ gauge theory with one chiral field discussed in Section \ref{sec: N=2 warm up} and in (\ref{Abelian SU/U duality}). For $k>1$ the theories are $\cN=1$.

\paragraph{Relevant deformations and vacua.} Let us compare the behavior of the two theories under the relevant deformation.

In Theory A (for $N\geq 2$) we take the deformed superpotential
\be
W = \frac m2 Q^\dag Q - \frac14 Q^\dag Q Q^\dag Q \;.
\ee
The F-term is $F_Q = q\big( m - |q|^2 \big)$. One finds the following vacuum structure:
\begin{itemize}
\item For $m>0$ there are two vacua.

One vacuum is at $q=0$ where $Q$ has mass $m$. Integrating it out we get the topological theory
$$
\cN=1 \;\; SU(N)_{k+ \frac N2} \quad\cong\quad SU(N)_k \;.
$$
We have indicated both the $\cN=1$ and the standard $\cN=0$ notation (see \cite{Witten:1999ds} and Appendix \ref{app: level-rank}).

The other vacuum is at $|q|^2 = m$, where the gauge symmetry is broken and the radial mode is massive. The breaking $SU(N) \to SU(N-1)$ eats $2N-1$ real bosonic modes, while the real radial mode is massive, therefore all modes of $q$ are massive. Taking $q = \big( \sqrt m\; ,\, 0 ,\dots, 0 \big)$ and analyzing the quadratic fermionic action, one finds that the modes $\psi_1$ and $\lambda_{11}$ acquire a mass, which does not affect the CS level of the unbroken group. The modes $\psi_a$ and $\lambda_{1a}$ ($a \neq 1$) give two modes of opposite mass, therefore the bare CS level of the unbroken gauge group is shifted by $-1$ (such a bare level was $k + N$) while the effective CS level is not shifted. Therefore we are left with
$$
\cN=1 \;\; SU(N-1)_{k + \frac N2- \frac12} \quad\cong\quad SU(N-1)_k \;.
$$

\item For $m<0$ there is a single vacuum at $q=0$ where $Q$ has mass $-|m|$. Integrating it out we get
$$
\cN=1 \;\; SU(N)_{k + \frac N2-1} \quad\cong\quad SU(N)_{k-1} \quad \text{(for $k\geq 1$)} \;.
$$
For $k=1$ this is a trivial gapped vacuum.
\end{itemize}

\noindent In Theory B (for $N\geq 2$) we take the deformed superpotential
\be
W = m H + H  P^\dag  P - \frac13 H^3 \;.
\ee
The F-terms are $F_H = m + | p|^2 - h^2$ and $F_P = 2h p$. One finds the following vacuum structure:
\begin{itemize}
\item For $m>0$ there are two vacua at $ p=0$ and $h = \mp \sqrt m$, where $ P$ has mass $\mp 2\sqrt m$. Also $H$ is massive, with mass $\pm 2 \sqrt m$ around its VEV. Integrating them out, in the vacuum with upper sign we get
$$
\cN=1 \;\; U(k)_{-N -\frac k2,\, - N} \quad\cong\quad U(k)_{-N} \;,
$$
while in the vacuum with lower sign we get
$$
\cN=1 \;\; U(k)_{-N -\frac k2 +1,\,  - N+1} \quad\cong\quad U(k)_{-N+1} \;.
$$

\item For $m<0$ there is a vacuum at $| p|^2 = |m|$ and $h = 0$, where the gauge symmetry is broken and the radial mode of $ p$ is massive. The singlet $H$ mixes with the radial part of $P$ around its VEV, giving two modes of opposite masses $\pm 2 \sqrt{|m|}$. Since $\chi_a$ and $\lambda_{1a}$ ($a\neq1$) give two modes of opposite mass, we are left with%
\footnote{The shift in the level $k'$ is only apparent: if we write $U(n)_{k, k + mn}$ we see that $m$ is not shifted.}
$$
\cN=1 \;\; U\big( k - 1 \big)_{-N - \frac k2 + \frac12,\, -N} \quad\cong\quad U(k-1)_{-N} \;.
$$
\end{itemize}
In all cases we find a perfect match between the two descriptions.

\

For $N=1$, Theory A is a free $\cN=2$ chiral multiplet. Under both positive and negative mass deformation, it gives a trivial gapped vacuum. The analysis of deformations of Theory B in (\ref{SU/U Theory B special N=1}) requires to keep into account the one-loop effects, as we did in Section~\ref{sec: N=2 warm up}. For $m>0$, $h$ gets negative VEV, $P$ acquires negative mass and integrating it out one gets a shift $\delta\cW = \frac 12 H^2$. This leads to $U(k)_{-1}$ CS theory, which has a trivial gapped vacuum. For $m<0$, we only find the Higgsed vacuum $|p|^2 = |m|$ leading to $U(k-1)_{-1}$ CS theory with a trivial gapped vacuum. The classical vacuum where $h$ gets a positive VEV is lifted quantum mechanically, because of the one-loop shift $\delta\cW = -\frac 12 H^2$.

\

Let us mention that, if we consider a theory as in (\ref{SU/U Theory A}) but with opposite sign of the superpotential, \ie{} $\cW = \frac14 Q^\dag Q Q^\dag Q$, then its vacuum structure is reproduced by a theory as in (\ref{SU/U Theory B}) but with superpotential $\cW = HP^\dag P + \frac\alpha2 H^2$ (with large positive $\alpha$) \ie{} with quadratic rather than cubic term in $H$. Then $H$ could be integrated out leading to a theory with no singlets and superpotential $\cW = -\frac1{2\alpha} P^\dag P P^\dag P$.

\subsection[$U/SU$ duality]{\matht{U/SU} duality}

The second family, that we call \matht{U/SU} and generalizes (\ref{basic N=1 duality}), is
\be
\ba{c} \text{$U(N)_{k+\frac N2 - \frac12,\, k - \frac12}$ with 1 $Q$} \\[.5em] \cW = -\frac14 Q^\dag Q Q^\dag Q \ea
\qquad\longleftrightarrow\qquad
\ba{c} \text{$SU(k)_{-N-\frac k2 + \frac12}$ with 1 $P$ and $H$} \\[.5em] \cW = HP^\dag P - \frac13 H^3 \;. \ea
\ee
We consider this duality in the range $N,k\geq 1$. The case $k=1$ is special because the RHS becomes the WZ model that we have studied in Section~\ref{sec: abelian}.

\paragraph{Relevant deformations and vacua.} Theory A (on the LHS) has the following vacuum structure:
\begin{itemize}
\item For $m>0$ there are two vacua. In the vacuum at $q=0$ the field $Q$ has mass $m$. Integrating it out gives the topological theory $U(N)_k$.

In the Higgsed vacuum at $|q|^2 = m$ we are left with $U(N-1)_k$.

Notice that for $k=1$ both gapped vacua are trivial, and for $N=1$ the second gapped vacuum is trivial.

\item For $m<0$, in the vacuum at $q=0$ the field $Q$ has mass $-|m|$. Integrating it out gives $\cN=1$ $U(N)_{k + \frac N2-1, k-1}$ CS theory. For $k>1$ this is the topological $U(N)_{k-1}$ CS theory, while for $k=1$ this is an $S^1$ free scalar together with a free fermion.
\end{itemize}
Theory B (on the RHS) has the following vacuum structure:
\begin{itemize}
\item For $m>0$ there are two vacua at $h= \mp\sqrt m$ where $P$ has mass $\mp 2\sqrt m$ and $H$ is massive as well. Integrating them out, in the vacuum with upper sign we get $SU(k)_{-N}$ while in the vacuum with lower sign we get $SU(k)_{-N+1}$.

\item For $m<0$ there is a Higgsed vacuum at $|p|^2 = |m|$, leading to $SU(k-1)_{-N}$ for $k>1$. When $k=1$, the global symmetry is broken and we get an $S^1$ Goldstone boson with a free fermion instead.
\end{itemize}
The two descriptions match.

\subsection[$U/U$ duality]{\matht{U/U} duality}

The third and fourth families, that we call \matht{U/U} and generalize (\ref{Abelian U/U duality +}) and (\ref{Abelian U/U duality -}), are
\be
\ba{c} \text{$U(N)_{k+\frac N2 - \frac12,\, k - \frac12 \pm N}$ with 1 $Q$} \\[.5em] \cW = -\frac14 Q^\dag Q Q^\dag Q \ea
\quad\longleftrightarrow\quad
\ba{c} \text{$U(k)_{-N-\frac k2 + \frac12,\, -N + \frac12 \mp k}$ with 1 $P$ and $H$} \\[.5em] \cW = HP^\dag P - \frac13 H^3 \;. \ea
\ee
We consider these dualities in the range $N, k \geq 1$.

\paragraph{Relevant deformations and vacua.} Theory A (on the LHS) has the following vacuum structure:
\begin{itemize}
\item For $m>0$ there are two vacua. In the vacuum at $q=0$ the field $Q$ has mass $m$. Integrating it out gives the topological theory $U(N)_{k,k\pm N}$.

In the Higgsed vacuum at $|q|^2 = m$ we are left with $U(N-1)_{k,k\pm(N-1)}$.

There are some special cases: when the second level is 0 we get an $S^1$ free scalar and a free fermion.

\item For $m<0$, in the vacuum at $q=0$ the field $Q$ has mass $-|m|$. Integrating it out gives $\cN=1$ $U(N)_{k + \frac N2-1, k-1 \pm N}$ CS theory. This is the topological $U(N)_{k-1,k-1\pm N}$ CS theory.
\end{itemize}
Theory B (on the RHS) has the following vacuum structure:
\begin{itemize}
\item For $m>0$ there are two vacua at $h= \mp\sqrt m$ where $P$ has mass $\mp 2\sqrt m$ and $H$ is massive as well. Integrating them out, in the vacuum with upper sign we get $U(k)_{-N,-N\pm k}$ while in the vacuum with lower sign we get $U(k)_{-N+1,-N+1\mp k}$.

\item For $m<0$ there is a Higgsed vacuum at $|p|^2 = |m|$, leading to $U(k-1)_{-N, -N\mp (k-1)}$.
\end{itemize}
The two descriptions match.

\subsection[Generalization to $N_f>1$]{Generalization to \matht{N_f>1}}
\label{sec: nflarge}

Our proposed dualities admit a natural generalization to the case with more than one flavor. Even if a detailed analysis of this case is beyond the scope of the present paper, let us state the conjecture and make a few comments. There are four families of dualities: the $SU/U$ duality
\be
\ba{c} \text{$SU(N)_{k + \frac N2 - \frac{N_f}2}$} \\[.7em] \text{with $N_f$ flavors $Q_i$} \\[.5em] \cW = - |Q|^4 \ea
\qquad\longleftrightarrow\qquad
\ba{c} \text{$U(k)_{-N-\frac k2 + \frac{N_f}2,\, -N + \frac{N_f}2}$} \\[.7em] \text{with $N_f$ flavors $P_j$, $N_f^2$ singlets $H_{ij}$} \\[.5em] \cW = H |P|^2 - H^3 \;, \ea
\ee
the $U/SU$ duality
\be
\ba{c} \text{$U(N)_{k + \frac N2 - \frac{N_f}2,\, k - \frac{N_f}2}$} \\[.7em] \text{with $N_f$ flavors $Q_i$} \\[.5em] \cW = - |Q|^4 \ea
\qquad\longleftrightarrow\qquad
\ba{c} \text{$SU(k)_{-N-\frac k2 + \frac{N_f}2}$} \\[.7em] \text{with $N_f$ flavors $P_j$, $N_f^2$ singlets $H_{ij}$} \\[.5em] \cW = H |P|^2 - H^3 \;, \ea
\ee
and the two $U/U$ dualities
\be
\ba{c} \text{$U(N)_{k + \frac N2 - \frac{N_f}2,\, k - \frac{N_f}2 \pm N}$} \\[.7em] \text{with $N_f$ flavors $Q_i$} \\[.5em] \cW = - |Q|^4 \ea
\qquad\longleftrightarrow\qquad
\ba{c} \text{$U(k)_{-N-\frac k2 + \frac{N_f}2,\, -N + \frac{N_f}2 \mp k}$} \\[.7em] \text{with $N_f$ flavors $P_j$, $N_f^2$ singlets $H_{ij}$} \\[.5em] \cW = H |P|^2 - H^3 \;. \ea
\ee
On both sides, the flavors are in the (complex) fundamental representation, while the $N_f$ singlets are real. The global symmetry is $U(N_f)/\bZ_N \rtimes \bZ_2^\cC$, where the second factor is charge conjugation. The singlets $H_{ij}$ transform in the adjoint plus singlet representation of the global $SU(N_f)$ symmetry factor, and the superpotentials are more complicated than in the $N_f=1$ case. On the LHS there are two possible terms (that we have schematically indicated by $|Q|^4$):
\be
\cW_\text{RHS} = - \big( \Tr Q^\dag Q \big)^2 - \Tr Q^\dag Q Q^\dag Q \;.
\ee
On the RHS there are, in principle, five possible terms (that we have schematically indicated by $H|P|^2 - H^3$):
\be
\cW_\text{LHS} = \Tr H P^\dag P + \Tr H \Tr P^\dag P - \big( \Tr H \big)^3 - \Tr H \Tr H^2 - \Tr H^3 \;.
\ee
We do not know the precise form of the superpotential nor the structure of the infrared fixed points of the two theories.

The dualities are expected to hold when $N_c \geq N_f$ on both sides, so we need $N \geq N_f$ and $k \geq N_f$. Outside this ranges there might be interesting quantum phases. We leave these issues to future work.

\acknowledgments{We are grateful to Sara Pasquetti for useful discussions. This work is supported in part by the MIUR-SIR grant RBSI1471GJ ``Quantum Field Theories at Strong Coupling: Exact Computations and Applications". S.B. is partly supported by the INFN Research Projects GAST and ST$\&$FI. Part of this project was completed at the workshop ``Superconformal Field Theories in 6 and Lower Dimensions" at the Tsinghua Sanya International Mathematics Forum.}


\appendix

\section{Cubic \matht{\CN=1} Wess-Zumino models in the \matht{\varepsilon}-expansion}
\label{loops}

We want to study perturbatively the cubic Wess-Zumino model with Lagrangian
\be
\cL = \int d^2\theta  \left( -\frac12 \sum\nolimits_{i=0}^K D_\a \Phi_i D^\a \Phi_i + \CW(\Phi_i) \right) \;.
\ee
Here $\CW(\Phi_i)$ is a cubic real function of the real superfields $\Phi_i(x,\theta)$:
\be
\CW = \frac{1}{6} \, g_{I,J,K} \, \Phi_{I}\Phi_J\Phi_K \;.
\ee
Each $\Phi$ expands as
\be
\Phi(\theta) = \phi + \theta \lambda + \theta^2 F_{\Phi}
\ee
where $\phi$ is a real scalar, $\lambda$ is a real (Majorana) fermion, and $F$ is an auxiliary field. 

One obstacle that we find in the study $\CN=1$ theories in the $\varepsilon$-expansion is the following: in $4d$, minimal fermions contain two copies of $3d$ real fermions. So the $4d$ loop computations in the literature will not directly provide the results for real $\CN=1$ superfields. One way around this obstacle is the following \cite{Fei:2016sgs}. This strategy should work for any Wess-Zumino model whose interacting Lagrangian is
\be
\partial_I \CW \, \partial^I \CW + \partial_I \partial_J \CW \, \lambda^I \lambda^J \;.
\ee

The fermions $\lambda$ are two-component Majorana fermions. We replace each $\lambda$ with a tower of $4p$ fields $\lambda_j$ (where $p$ is an arbitrary integer) and modify the fermionic part of the previous Lagrangian as follows:
\be
\partial_I \CW \, \partial^I \CW +   \partial_I \partial_J \CW \, \sum_{j=1}^{4p} \lambda^I_j \lambda^J_j \;.
\ee

We do not change the number of scalars nor the quartic scalar interactions.  At this point we combine the $4p$ $\lambda$'s into $p$ complex Dirac four-component fermions. We obtain a Gross-Neveu-Yukawa model with global symmetry $SU(p)$ that exists for any $d \leq 4$. At $d=3$ the global symmetry enhances to $SO(4p)$. For this extended model we can use the existing results for loop computations present in the literature. They are the beta-functions of the quartic and Yukawa couplings, and also the scaling dimensions of the fields $\phi_I$, $\lambda_I$ and $\phi_I \phi_J$. At some point in the computation we can set $p=1/4$ and infer the results for our $3d$ real fermions, hence for our cubic Wess-Zumino model of interest. This strategy was implemented in \cite{Fei:2016sgs}.

Our interest in this paper lies in the following cubic Wess-Zumino model,\footnote{A similar model, but with $\CN=2$ supersymmetry, was studied in \cite{Chester:2015qca}. It was found that the coupling $g_3$ flows to $0$ at the IR fixed point, leading to accidental IR symmetries. In our case with $\CN=1$ supersymmetry we find that both $g_2$ and $g_3$ are non-vanishing in the IR fixed point. Even starting from $g_3=0$ in the UV, the term $\Phi_0^3$ would be generated in the IR.} with two independent couplings:
\be
\CW= \frac{g_2}{2} \Phi_0 \sum_{i=1}^K\Phi_i^2 + \frac{g_3}{6} \Phi_0^3 \;.
\ee
The global symmetries are $O(K)$ and parity, which forbid other cubic terms $\Phi_{i>0}^3$ and $\Phi_{i>0}\Phi_0^2$ to be generated. Under parity the $\Phi_i$'s are even and $\Phi_0$ is odd. Notice also that only the relative sign between $g_2$ and $g_3$ is physical; we will find that $g_2$ and $g_3$ have opposite sign at the RG fixed point.

We compute the beta-functions for Yukawa couplings (and hence for our SUSY couplings) using eqn. (7.2) of \cite{Jack:1990eb}.
In terms of the two SUSY couplings $g_2,g_3$, the beta-functions and scaling dimensions at one-loop are
\bea
\beta_{g_3} &= -\frac{\varepsilon}{2}g_3+ \frac{1}{2(4\pi)^2} \left( 7 g_3^3 + 3K g_3 g_2^2 + 4 K g_2^3 \right) \\
\beta_{g_2} &= -\frac{\varepsilon}{2}g_2+\frac{g_2}{2(4\pi)^2}\left( g_3^2 + 4 g_3 g_2 + ( K + 8 ) g_2^2 \right) \;.
\eea
We compute the scaling dimension of the fundamental fields in the Wess-Zumino model using eqn. (A.3) of \cite{Fei:2016sgs}. At one-loop they are:
\be
\Delta[\Phi_0] =  \frac{2-\varepsilon}{2} +   \frac{g_3^2 + K g_2^2}{2(4\pi)^2} \;,\qquad\qquad
\Delta[\Phi_i] =  \frac{2-\varepsilon}{2} +   \frac{2 g_2^2}{2(4\pi)^2} \;.
\ee
From eqns. (A.4) and (A.5) of \cite{Fei:2016sgs} we can obtain the mixing matrix for the operators quadratic in the fundamental fields. The $\frac{(K+1)(K+2)}{2}$ quadratic operators transform in the vector, symmetric traceless ($\CO_{s.t.}$) and two singlets representations of $SO(K)$.

The operators $\Phi_0 \Phi_i$ transform as a vector of $SO(K)$, so they do not mix with the other quadratic operators. At one-loop their scaling dimension is given by
\be
\Delta[\Phi_0 \Phi_i] =  2-\varepsilon +  \frac{g_3^2 + (K+10) g_2^2+ 4 g_2 g_3}{2(4\pi)^2} \;.
\ee
Notice that, upon using $\b_{g_2}=0$, the relation $\Delta[\Phi_0 \Phi_i] = \Delta[\Phi_i] +1$ is satisfied. This is consistent with the fact that $\Phi_0\Phi_i$ is a SUSY descendant of $\Phi_i$ at the fixed point.

We write down the mixing matrix for the $K+1$ operators $\Phi_0^2,\Phi_1^2,\Phi_2^2, \ldots$, at one-loop:
\begin{multline}
\Delta[\Phi_0^2,\Phi_i^2] =  (2-\varepsilon )\mathbb{I}_{K \times K} + {} \\
{} + \frac{1}{(4\pi)^2} 
           \left(\ba{ccccc} K g_2^2\!+\! \,\,4 g_3^2 &\,\,  g_2 (2g_2\!+\!g_3) \,\, &\,\,  g_2 (2g_2\!+\!g_3) \,\, &\,\,  g_2 (2g_2\!+\!g_3)&\,\,  \dots  \\
                            g_2 (2g_2\!+\!g_3)    & 5 g_2^2  & g_2^2 & g_2^2& \ldots \\
                              g_2 (2g_2\!+\!g_3)    & g_2^2   &5 g_2^2& g_2^2 & \ldots \\
                       g_2 (2g_2\!+\!g_3)   & g_2^2 & g_2^2 & 5 g_2^2 & \ldots \\
                                \vdots     & \vdots & \vdots & \vdots & \ddots   
                               \ea \right)
\end{multline}
From its eigenvalues we can read off the scaling dimensions of the singlet operator $\CO_s \simeq \Phi_0^2 + \sum \Phi_i^2$ (at one-loop), the singlet $g_2 \Phi_0^2 + g_3 \sum \Phi_i^2$ (which is a SUSY descendant of $\Phi_0$), and the symmetric traceless operator $\CO_{s.t.}$. We report the one-loop results in Table~\ref{table}. 

\begin{table}\label{table}
\centering
\begin{tabular}{|c|cccccccc|}\hline
                                                                   & $K=0$ & $K=1$ & $K=2$ & $K=4$ & $K=6$  & $K=10$  & $K=20$ & $K=50$ \\ \hline
$\frac{g_2}{4\pi\sqrt{\varepsilon}}$  & &       $1/\sqrt{6}$ &  $0.382$  &    $0.340$&$0.308$&  $0.264$&  $0.203$& $0.136$ \\ \hline
$-\frac{g_3}{4\pi\sqrt{\varepsilon}}$&$1/\sqrt{7}$&$1/\sqrt{6}$&$0.414$&$0.407$&$0.393$&  $0.364$&  $0.311$& $0.233$ \\ \hline
$\Delta[\Phi_0]$                                   & $4/7$ &    $2/3$ &   $0.732$  &  $0.814$&$0.862$&  $0.914$& $0.961$& $0.990$ \\ \hline
$\Delta[\Phi_i]$                          &               &    $2/3$ &   $0.646$    &    $0.616$  &$0.595$&  $0.570$& $0.541$& $0.518$ \\ \hline
$\!\!\Delta[\CO_s]\!\!$                  & $11/7$ &    $2$ &   $2.124$    &    $2.236$&   $2.275$   & $2.285$&$2.243$ & $2.151$ \\ \hline
$\!\!\!\Delta[\CO_{s.t.}]\!\!\!$           &      &    $5/3$ &   $1.585$    &    $1.463$&   $1.380$    &  $1.278$& $1.165$ & $1.074$ \\ \hline
\end{tabular}
\caption{The Wess-Zumino model $\CW= \frac{g_2}{2} \Phi_0 \sum_{i=1}^K\Phi_i^2 + \frac{g_3}{6} \Phi_0^3$ at one-loop: coupling at the RG fixed point and scaling dimensions of the elementary and quadratic operators. We are interested in even $K>0$, but we also consider the cases $K=0$ and $K=1$ in order to compare with existing results in the literature.}
\end{table}

Let us make a few comments about the different values of $K$.

\paragraph{\underline{$K=0$.}} This case is dubbed the supersymmetric Ising model, $\CW=\Phi_0^3$. The scaling dimension at two loops is
\be
\Delta[\Phi_0] = 1- \frac{3}{7}\varepsilon +\frac{1}{49}\varepsilon^2 + O(\varepsilon^3) \;\;\simeq\; 0.59
\ee
in agreement with  \cite{Fei:2016sgs} and with the numerical boostrap results of \cite{Iliesiu:2015qra} $\Delta \simeq 0.582$. Since $\phi_0^2$ is a descendant of $\phi_0$, it follows $\Delta[\Phi_0^2] = \Delta[\Phi_0] + 1$.

\paragraph{\underline{$K=1$.}} In this case the model has an emergent $\CN=2$ supersymmetry, at the critical point $g_2=-g_3$ (we checked this statement at two-loops), so the theory is the $\CN=2$ $\Phi^3$ Wess-zumino model, with $\Phi = \Phi_0 + i \Phi_1$:
\be
\CW = g_3 \, \Big( -\frac{1}{2} \Phi_0 \Phi_1^2 + \frac{1}{6} \Phi_0^3 \Big) =\frac{g_3}{12} (\Phi_0 + i \Phi_1)^3 + c.c.
\ee
The scaling dimensions of the elementary fields are one-loop exact: $\Delta[\Phi_0] = \Delta[\Phi_1] = \frac{2}{3}$. Then $\CO_{s.t.}$ is a SUSY descendant so $\Delta[\CO_{s.t.}]=\frac{5}{3}$. On the other hand, the operator $\CO_s=\Phi_0^2+\Phi_1^2 \simeq \Phi \Phi^\dagger$ gets corrections beyond one-loop:
\be
\Delta[\cO_s] =  2 - \frac{1}{3}\varepsilon^2 +\frac{1+12\zeta(3)}{18} \varepsilon^3 + O(\varepsilon^4) \;.
\ee
Its precise scaling dimension is $\simeq 1.91$ (obtained by resuming three or four loops in the $\varepsilon$-expansion \cite{Zerf:2016fti, Fei:2016sgs, Baggio:2017mas} or by numerical bootstrap \cite{Bashkirov:2013vya, Bobev:2015vsa, Bobev:2015jxa}). Notice that in this case the two-loop result is not very close to the precise value, because the one-loop correction is vanishing.

\paragraph{\underline{$K=2$.}}
The scaling dimension of the supersymmetric primary $O(2)$-singlet quadratic operator $\Phi_0^2+0.54187\sum \Phi_i^2$ is greater than $2$, so the deformation induced by this ``mass term'' is irrelevant. This is consistent with our conjectured duality. In Section \ref{2loop} we report the results of a two-loop computation. The Pade resumed value of $\Delta[\cO_s]$ is still larger than $2$, giving further support to the statement.

The scaling dimension of the operators $\CO_{s.t.}$ in the symmetric traceless is instead smaller than $2$, so the mass deformations $\Phi_1\Phi_2$, $\Phi_2^2-\Phi_1^2$ are relevant. In the dual gauge theory, assuming the duality, this means that the monopole square deformations $\M^{\pm 2}$ are relevant.

\paragraph{\underline{$K>2$.}}
The main point we want to emphasize is that for any $K>1$, the one-loop scaling dimension of the singlet quadratic operator $\cO_s = \Phi_0^2 \tp  \sum \Phi_i^2$ is greater than 2. The maximum in $K$ is reached at $K=9$ while at large $K$, $\Delta[\cO_s] \rightarrow 2^+$. But at large $K$ higher-loop corrections are suppressed, so the one-loop result is reliable.

For $K$ even, we can gauge an $SU(\frac{K}{2})$ or a $U(\frac{K}{2})$ subgroup of $O(K)$. If the Chern-Simon coefficient is large enough, the property $\Delta[\cO_s]>2$ will not be spoiled by the gauging. This is a consistency check of our proposed dualities. Notice also that if the Chern-Simons coefficient vanishes, the term $\Phi_0^2$ is forbidden by parity invariance.

\section{Level-rank dualities in \matht{\cN=1} notation}
\label{app: level-rank}

First of all we need the following general facts. $SU(N)_k$ requires $k \in \bZ$. If we integrate out a (real) fermion in the adjoint representation, we shift $k$ by $\pm \frac N2$. Therefore
\be
\text{$\cN=1$ $SU(N)_k$ requires $k - \tfrac N2 \in \bZ$} \;.
\ee
We have
\be
\cN=1 \;\; SU(N)_k \quad\xrightarrow{m_\lambda}\quad \begin{cases} SU(N)_{k+ \frac N2} &\text{for } m_\lambda>0 \\ SU(N)_{k-\frac N2} &\text{for } m_\lambda<0 \;. \end{cases}
\ee
In our conventions, the sign of the fermion mass at the $\cN=1$ point is opposite to the sign of $k$.

The theory $U(N)_{k,k'}$ requires $k = k' \mod{N}$. We can write
\be
U(N)_{k,\, k + MN} = \frac{SU(N)_k \times U(1)_{MN^2}}{\bZ_N} \qquad\qquad M \in \bZ \;.
\ee
When integrating out a (real) fermion in the adjoint representation, we shift $k$ by $\pm \frac N2$ while $k'$ does not shift. Therefore
\be
\text{$\cN=1$ $U(N)_{k,k'}$ requires $k - \tfrac N2 \in \bZ$ and $k' = k - \tfrac N2 \mod{N}$} \;.
\ee
In particular $k'$ cannot be equal to $k$.

We recall the level-rank dualities of spin-TQFTs:
\bea
SU(N)_k &\qquad\longleftrightarrow\qquad U(k)_{-N} \\
U(N)_{k, k\pm N} &\qquad\longleftrightarrow\qquad U(k)_{-N, - N \mp k}
\eea
for $N>0$, $k>0$. Assuming $N,k$ positive, we can write the following $\cN=1$ level-rank dualities:
\bea
\cN=1 \;\; SU(N)_{k + \frac N2} &\quad\longleftrightarrow\quad \cN=1 \;\; U(k)_{- N - \frac k2,\, -N} \\
\cN=1 \;\; U(N)_{k + \frac N2,\, k \pm N} &\quad\longleftrightarrow\quad \cN=1 \;\; U(k)_{-N - \frac k2,\, -N \mp k} \;.
\eea
All these dualities are valid for $N,k>0$.

\bibliographystyle{ytphys}
\bibliography{Non_SUSY_dualities}
\end{document}